\documentclass[aps,prb,twocolumn,groupedaddress,showpacs,superscriptaddress]{revtex4}

\usepackage{graphics,graphicx}
\usepackage{enumitem}
\graphicspath{{pict/}{}}

\newcommand\lsect[1]{\protect\label{sect:#1}}

\begin{document}

\title{Bloch-mode analysis for retrieving effective parameters of metamaterials}

\author{Andrei Andryieuski}
\email[]{andra@fotonik.dtu.dk}
\affiliation{DTU Fotonik - Department of Photonics Engineering, Technical University of Denmark, {\O}rsteds pl. 343, DK-2800 Kongens Lyngby, Denmark}

\author{Sangwoo Ha}
\author{Andrey A. Sukhorukov}
\author{Yuri S. Kivshar}
\affiliation{Nonlinear Physics Centre and Centre for Ultrahigh-bandwidth Devices for Optical Systems (CUDOS), Research School of Physics and Engineering, Australian National University, Canberra, ACT 0200, Australia}

\author{Andrei V. Lavrinenko}
\affiliation{DTU Fotonik - Department of Photonics Engineering, Technical University of Denmark, {\O}rsteds pl. 343, DK-2800 Kongens Lyngby, Denmark}
\date{\today}

\begin{abstract}
We introduce a new approach for retrieving effective parameters of metamaterials based on the Bloch-mode analysis of quasi-periodic composite structures. We demonstrate that, in the case of single-mode propagation, a complex effective refractive index can be assigned to the structure, being restored by our method with a high accuracy. We employ both surface and volume averaging of the electromagnetic fields of the dominating (fundamental) Bloch modes to determine the Bloch and wave impedances, respectively. We discuss how this method works for several characteristic examples, and demonstrate that this approach can be useful for retrieval of both material and wave effective parameters of a broad range of metamaterials.
\end{abstract}


\maketitle

\section{Introduction} \lsect{intro}

The study of artificially structured metamaterials (MMs) attracts attention of scientists and engineers due to their unprecedented electromagnetic properties. Negative refractive index, very large or near zero values of both permittivity and permeability, giant optical activity~-- these are just a few examples of the properties which MMs can provide~\cite{MMHandbook}. As was established, it is convenient to describe the MM properties by employing the concept of effective parameters (EPs), such as refractive index $n$, impedance $z$, permittivity $\varepsilon$ and permeability  $\mu$, provided that these EPs can be introduced~\cite{Simovski2009_optspectrosc}. The EPs simplify significantly the description of the MM properties, including the propagation of electromagnetic waves inside a MM slab and their reflection and transmission at the MM boundaries.

The state-of-the-art of homogenization infers that retrieved EPs  are of two types~\cite{Menzel2008_inclined,Simovski2009_optspectrosc,Simovski2007_analyt}:

(i) Material (or local) effective parameters (MEP) $\varepsilon_M$ and $\mu_M$. They give the relation of the field vectors $\textbf{D}=\varepsilon_M\varepsilon_0\textbf{E}$ and $\textbf{B}=\mu_M\mu_0\textbf{H}$. The material effective parameters show the evolution of the wave inside a metamaterial. Material EPs depend only on the properties of the material (we do not consider here the problem of the Drude transition layers\cite{Simovski2009_optspectrosc}). Specifically, material EPs are important, for example, for the superlens performance of the slab with negative refractive index\cite{Pendry2000}. The relations to the refractive index $n$ and wave impedance $z_W$ are:
        \begin{equation} \label{EQ1}
            n=\sqrt{\varepsilon_M\mu_M},
        \end{equation}
        \begin{equation} \label{EQ2}
            z_W=\sqrt{\mu_M/\varepsilon_M}.
        \end{equation}

(ii) Wave (or nonlocal) effective parameters (WEP) $\varepsilon_W$ and $\mu_W$. They are usually restored from the reflection and transmission coefficients of a MM slab~\cite{Smith2002_SM} being assigned as the parameters of the corresponding homogenous slab. Sometimes this approach leads to violation of \emph{locality} conditions, and this situation was actively discussed in the literature~\cite{Koschny2003_antiresonance, Depine2004_antiresonance, Efros2004_antiresonance, Koschny2004_antiresonance, Simovski2007, Menzel2008_inclined, Simovski2009_optspectrosc, Simovski2007_analyt}. The WEP may allow one solving the scattering problem (reflection/transmission determination) for a MM slab of another thickness. They often depend on the thickness of the MM slab (in terms of the number of unit cells, see e.g. Ref.~[\onlinecite{Zhou2009}]), with only rare exceptions~\cite{Andryieuski2009}.

For a homogeneous medium with the structural element characteristic size $a$, which is much less than the wavelength $\lambda$, the material and wave EPs are the same. However, in many practical cases MM's unit cell is only $a\sim\lambda/10-\lambda/4$ and material and wave parameters are not equivalent to each other \cite{Simovski2007_analyt}. It is obvious that the reflection from a MM slab should depend on whether the MM slab termination coincides with the border or with another cross-section somewhere in the middle of the unit cell, so the wave EPs depend on the MM opening cross-section.

The knowledge of the WEP and MEP is needed for development of metamaterial based devices. This would be desirable to obtain both sets of EPs within a similar simple calculation procedure. The importance of the EPs restoration is emphasized by a variety of the existing retrieval methods, which are summarized in the Table \ref{Table}.

\begin{center}
\begin{table}
\caption{\label{Table} Comparison of the EPs restoration methods}
    \begin{tabular}{ | p{4cm} | p{4cm} |}
    \hline
    Method and References. & Effective parameters type and comments\\
    \hline
    Reflection-transmission (Nicholson-Ross-Weir(NRW)) \cite{Smith2002_SM,ChenX2004_robust,Menzel2008_inclined,Cook2009_KK} & WEP: Scalar, restored for normal or inclined incidence.\\
    \hline
    Wave propagation \cite{Popa2005_fields,Andryieuski2009_WPRM} & WEP: Scalar, restored for normal incidence.\\
    \hline
    Field averaging \cite{Lerat2006_fieldsumm, Smith2006_fieldaver, Acher2007_fieldsumm, Pors2011, Tsukerman2011} & MEP: Scalar or tensor.\\
    \hline
    Analytical and semi-analytical \cite{Chern2009_analyt,Morits2010_analyt,Petschulat2008_multipole,Simovski2007_analyt,Zhukovsky_tt} & MEP: Tensor.\\
    \hline
    Single interface scattering \cite{Yang2010_singlerefl} & WEP: Scalar, restored for normal incidence.\\
    \hline
    Non-local dielectric function \cite{Costa2011,Silveirinha2011,Silveirinha2007a,Silveirinha2007,Silveirinha2009,Costa2009_FDFDnonlocal,Chebykin2011} & MEP: Nonlocal dielectric function, tensor.\\
    \hline
    Current-driven \cite{Fietz2010, Fietz2009} & MEP: Tensor.\\
    \hline
    Quasi-mode \cite{Sun2009_quasimode} & WEP: Scalar, restored for normal incidence.\\
    \hline
    \end{tabular}
    \end{table}
\end{center}

This paper aims to introduce and discuss in detail an approach described in [\onlinecite{Simovski2009_optspectrosc, Simovski2011}] for retrieving the wave and material effective parameters. First, we calculate the dispersion bands of the long enough periodic media by employing the high-resolution spectral analysis method~\cite{Ha:2009-3776:OL,Ha2011,Sukhorukov:2009-3716:OE}. This method is developed for periodic structures (in fact, quasi-periodic, taking into account a finite size of the structure) composed of arbitrary unit cells. After defining the dispersion of the dominating Bloch modes, we introduce a complex refractive index which can be attributed to the effective parameters of the metamaterial with a high accuracy.

Next, we introduce an effective impedance. Following Refs.~[\onlinecite{Simovski2009_optspectrosc, Simovski2011}], we apply the volume or surface averaging of the electric and magnetic fields of the dominating Bloch mode, which leads to the wave or Bloch (input) impedance EPs retrieval. Having both refractive index and impedance, we restore effective permittivity and permeability accordingly to Eqs.~(\ref{EQ1}) and (\ref{EQ2}), which will be either MEP or WEP, respectively. However, in contrast to the refractive index retrieving, the wave impedance retrieving procedure may encounter problems especially in application to MMs with the negative refractive index. Caution should be paid to the fields computed via direct numerical solution of Maxwell's equation by Maxwell's solvers. For example, in the CST Microwave Studio, which we used, the returned magnetic field calculated on a grid is magnetic induction $\textbf{b}/\mu_{0}$ and not magnetic strength \textbf{h}. Ignoring this fact when restoring the impedance from the electric and magnetic fields ratio can cause the real part of impedance becoming negative in the region of the negative refractive index, and correspondingly the negative energy flux is obtained. Such flux behavior is connected with its definition through the \textbf{H} field, the fact that was emphasized by Silverinha et al\cite{Silveirinha2009,Costa2011}.

The paper is organized as follows. In Sec.~\ref{Method} we formulate the general concept and technical details of our approach. The successful MEP retrieving examples in the case of homogeneous media and different types of composite MMs are summarized in Sec.~\ref{CaseStudies}. In Sec.~\ref{CaseStudies}, we also present the examples when the wave impedance retrieval leads to incorrect interpretation of EPs and, as a consequence, it connects impedance with the energy flux with wrong flux direction. Finally, in the concluding Sec.~\ref{Conclusions} we discuss both advantages and constraints of the novel approach introduced here.

\section{General approach} \label{Method}

The dispersion analysis is based on the Bloch modes expansion of the field propagating inside a MM slab. We simulate the field propagation by the commercial CST Microwave Studio software \cite{CST} with the finite-integrals Maxwell solver.

\begin{figure}
    \includegraphics[width=\columnwidth,keepaspectratio]{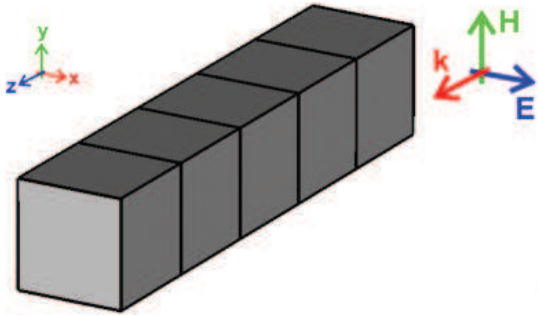}%
    \caption{\label{Simulation}(Color online). Simulation configuration. Wave is normally incident from vacuum. Wave propagation and metamaterial stacking direction is along $z$-axis. Electric field of the plane wave is polarized along x-axis.}
\end{figure}

We excite the MM slab, which consists of the periodically arranged unit cells of the period $\mathbf{a}=(a_x,a_y,a_z)$, with a plane wave propagating along the $z-$axis and electric field polarized along $x-$axis (see Fig.~\ref{Simulation}). In principle, the slab may be arbitrarily thick, but not less than 3-4 MM monolayers for that we can neglect the so-called Drude transition layers~\cite{Simovski2009_optspectrosc}.

We use perfect electric, perfect magnetic and open boundary conditions for the $x-$, $y-$ and $z-$ boundaries respectively and the time-domain solver in calculations. A broadband Gaussian pulse is used as a field source. Only one simulation is needed for the whole spectrum calculation. The fields on different frequencies are calculated through the Fourier transformations from the time-dependent signals collected with 3D field monitors.

Let us consider the plane wave normally incident from vacuum onto the MM slab. Its electric $E_{v}=E_{v0}\exp(ik_0 z)$ and magnetic $H_{v}=H_{v0}\exp(ik_0z)$ fields are connected via the impedance of the free space, $Z_0=E_{v0}/H_{v0}=\sqrt{\mu_0/\varepsilon_0} \approx 120 \pi$ Ohm. Here $k_0=\omega/c$ - is the wavenumber of the free space, and we assume the $\exp(-i\omega t)$ time dependence.

In general case, several Bloch modes~\cite{Yeh:1988:OpticalWaves, Joannopoulos:1995:PhotonicCrystals, Russell:1995-585:ConfinedElectrons, Mortensen2010_blochmodes} may be excited in the slab for each frequency $\omega$, so the overall field may be represented as a sum
\begin{equation} \label{BlochSumE}
    E(\mathbf{r}) = \sum_{m=1}^M E_m(\mathbf{r}),
\end{equation}
\begin{equation} \label{BlochSumH}
    H(\mathbf{r}) = \sum_{m=1}^M H_m(\mathbf{r}),
\end{equation}
where $m$ is the Bloch mode number, $M$ is the total number of excited modes, and $\mathbf{r} = (x,y,z)$. In the desirable case of local quasi-homogeneous MM there are only two modes in the slab: one forward and one backward propagating. A larger number of modes may be excited in the case of MM with strong spatial dispersion~\cite{Simovski2009_optspectrosc}.

The field profiles of Bloch modes can be represented as~\cite{Simovski2009_optspectrosc,Yeh:1988:OpticalWaves, Joannopoulos:1995:PhotonicCrystals, Russell:1995-585:ConfinedElectrons}
\begin{equation} \label{BlochE}
    E_m(\mathbf{r}) = \left[E_{m,0}(\mathbf{r}_\bot) + \sum_{p\neq 0}E_{m,p}(\mathbf{r}_\bot) e^{i G p z}\right] e^{i K_m z},
\end{equation}
\begin{equation} \label{BlochH}
    H_m(\mathbf{r}) = \left[H_{m,0}(\mathbf{r}_\bot) + \sum_{p\neq 0}H_{m,p}(\mathbf{r}_\bot) e^{i G p z}\right] e^{i K_m z},
\end{equation}
where $K_m$ is the Bloch wavenumber, $G = 2\pi /a_z$, $p$ is an integer number.
We note that the field representation in Eqs.~(\ref{BlochSumE}),~(\ref{BlochSumH}) is invariant with respect to a transformation $K_m \rightarrow K_m + G p'$ and $E_{m,p} \rightarrow E_{m,p+p'}$ for an arbitrary integer $p'$. Accordingly, we can always select the value of $K_m$ such that $E_{m,0}$ is the largest harmonic amplitude, and we use this convention in the following.

The key feature of the high-resolution spectral analysis method~\cite{Sukhorukov:2009-3716:OE, Ha:2009-3776:OL} is  decomposition of the total field obtained in simulations into a sum of Bloch modes, effectively inverting Eqs.~(\ref{BlochSumE}),~(\ref{BlochSumH}). The only prior information required for the application of this method is the number of strongest Bloch modes excited in the structure ($M$). Then, through specialized numerical fitting described in Refs.~[\onlinecite{Sukhorukov:2009-3716:OE, Ha:2009-3776:OL}] we extract wavenumbers $K_m$ and field profiles $E_m(\mathbf{r}), H_m(\mathbf{r})$ of all forward and backward propagating Bloch modes at each frequency $\omega$. By monitoring the accuracy of such decomposition in terms of field matching, we check whether other ignored Bloch modes have significant excitation amplitudes, and if this is a case we increase the number $M$ to take more modes into account and repeat the whole decomposition procedure.

It is an important advantage of our approach that the standing wave, which is usually formed inside the slab due to the multiple reflections from the boundaries and brings the restrictions to the conventional wave propagation retrieval method~\cite{Andryieuski2009_WPRM}, is not an issue in the present case, since we can separate forward and backward propagating Bloch modes. In the following, we denote the field profiles of the dominant forward and backward waves as
\begin{equation} \label{EH_pm}
  \{E,H\}_+ \equiv \{E,H\}_{m_+}, \quad \{E,H\}_- \equiv \{E,H\}_{m_-},
\end{equation}
and the corresponding wavenumbers
\begin{equation} \label{K_pm}
  K_+ \equiv K_{m_+}, \quad K_- \equiv K_{m_-},
\end{equation}
where $m_+$ and  $m_-$ are the numbers of the dominant forward and backward Bloch modes, respectively.

If several Bloch modes are excited and propagate in a MM, such composite cannot be homogenized and no meaningful EPs can be introduced. The homogeneity of MM and the influence of the higher-order Bloch modes have been discussed extensively in the Refs.~[\onlinecite{Andryieuski2009,Rockstuhl2008_fishnet_BM,Menzel2010}]. However, if only one forward mode can be distinguished by the lowest damping, we can count it as the dominating one and neglect the presence of the higher-order modes. As a rule it is the fundamental Bloch mode. The numerical criterion of homogeneity from the Bloch modes point of view was formulated in Ref.~[\onlinecite{Andryieuski2010}]. Another possibility to check the single mode regime is to calculate the mismatch $\delta$ of the restored sum of forward and backward propagating fundamental mode fields, $E_f=E_+ + E_-$, and the original field $E$ taken directly from numerical simulations:
\begin{equation}
    \delta= \frac{\int |E-E_f|^2dxdydz}{\int |E|^2dxdydz},
\end{equation}
where integration is performed over the computation domain. In all the case studies presented below the mismatch $\delta$ is below 1.5\%. So, in this manuscript we consider the MMs that have a dominant fundamental mode, and  the higher-order Bloch modes can be neglected. According to the concept of homogenization, we aim to find effective parameters for an equivalent homogeneous medium, where the wave propagation would be essentially the same as in the periodic structure. After determining the propagation constant $K_{+}$ of the fundamental mode we assign our structured material with the effective refractive index $n=K_{+}/k_0$.

The second part in restoration is connected with the effective impedance. We use the fields $E_+, H_+$ of the fundamental Bloch mode in the both Bloch $z_B$ and wave $z_W$ impedances restoration. First, we perform fields surface averaging at the $(x,y)$ cross-section of the simulated slab:
\begin{equation} \label{E_SA}
    E_{\rm SA}(z)=\int_{S}E_+(x,y,z)dxdy/a_xa_y,
\end{equation}
\begin{equation} \label{H_SA}
    H_{\rm SA}(z)=\int_{S}H_+(x,y,z)dxdy/a_xa_y.
\end{equation}
Taking the values of the fields $E_{{\rm SA},j}=E_{\rm SA}(z_j), H_{{\rm SA},j}=H_{\rm SA}(z_j)$ at the unit cell borders $z_j=j a_z$, where $j$ is an integer number, we determine the Bloch impedance~\cite{Simovski2007_analyt}:
\begin{equation}
\label{z_B}
    z_B=\frac{E_{{\rm SA},j}}{Z_0 H_{{\rm SA},j}}.
\end{equation}
Note that Bloch impedance $z_B$ does not depend on $j$, which can be checked by substituting Eqs.~(\ref{BlochE}) and~(\ref{BlochH}) into Eqs.~(\ref{E_SA}) and~(\ref{H_SA}),

In order to restore wave impedance ($z_W$), we need to calculate the volume averaged fields~\cite{Simovski2009_optspectrosc} $E_{\rm VA}$ and $H_{\rm VA}$,
\begin{equation} \label{z_W}
    z_W=\frac{E_{\rm VA}}{Z_0 H_{\rm VA}}.
\end{equation}
Since the wavenumbers in the periodically structured and equivalent homogeneous media are equal, we need to establish the correspondence of the field amplitudes in front of the common $\exp(i K_{+} z)$ multiplier. Accordingly, we define the volume-averaged fields by performing integration over a single unit cell with the multiplier $\exp(-i K_{+} z)$ to cancel the phase evolution:
\begin{equation} \label{E_VA}
    E_{\rm VA}=\int_{z_b}^{z_b+a_z} E_{\rm SA}(z) \exp(-i K_{+} z) dz/a_z,
\end{equation}
\begin{equation} \label{H_VA}
    H_{\rm VA}=\int_{z_b}^{z_b+a_z} H_{\rm SA}(z) \exp(-i K_{+} z) dz/a_z ,
\end{equation}
where $z_b$ is an arbitrary location inside the structure. We can also express the averaged fields through the harmonic amplitudes by substituting Eqs.~(\ref{BlochE}) and~(\ref{BlochH}) into Eqs.~(\ref{E_VA}) and~(\ref{H_VA}),
\begin{equation} \label{E_VA_F}
    E_{\rm VA}=\int_{S} E_{m+,0}(x,y) dxdy/a_x a_y,
\end{equation}
\begin{equation} \label{H_VA_F}
    H_{\rm VA}=\int_{S} H_{m+,0}(x,y) dxdy/a_x a_y.
\end{equation}
We see that the volume-averaged fields do not depend on $z_b$, as their values are defined through the dominant Bloch-wave harmonic amplitude which is $z$-independent.

For the extraction of $\textbf{E}$ and $\textbf{H}$ fields from the CST Microwave Studio simulations we use electric and magnetic field monitors. However, the raw microscopic magnetic field that CST returns is not  $\textbf{h}(\mathbf{r})$, but rather $\textbf{b}(\mathbf{r})/\mu_0$ as a straightforward solution of microscopic Maxwell's equations. As this was shown by M.~Silveirinha et al. \cite{Silveirinha2009, Costa2011}, the employment of the volume averaged magnetic induction $\textbf{B}_{VA}(\mathbf{r})$ instead of $\textbf{H}_{VA}(\mathbf{r})$ can give an incorrect direction of the Poynting vector for negative index metamaterials.

For the correct determination of the volume averaged magnetic field we employ the definition
\begin{equation}
   \textbf{H}_{VA}=\frac{\textbf{B}_{VA}}{\mu_0}-\textbf{M}_{VA},
\end{equation}
\begin{equation}
      \textbf{M}_{VA}= \frac{\int_{V}(\mathbf{r}\times\mathbf{J})dV}{2V},
\end{equation}%
where $\textbf{M}_{VA}$ - is the volume averaged magnetization vector and $\mathbf{J}$ is the current density. In principle the magnetization can be calculated by a numerical integration routine directly from the definition. However, we choose another, more elegant, approach following the findings of Silveirinha for the
transverse-averaged magnetic fields\cite{Silveirinha2007a}. First, we decompose $\textbf{M}_{VA}$ into two parts: along the direction of propagation (unit vector $\hat{\textbf{u}}_{z}$) and orthogonal to it
 \begin{equation}
   \textbf{H}_{VA}=\frac{\textbf{B}_{VA}}{\mu_0}-(\textbf{M}_{VA}\cdot\hat{\textbf{u}}_{z})\hat{\textbf{u}}_{z}+
   \hat{\textbf{u}}_{z}\times(\hat{\textbf{u}}_{z}\times\textbf{M}_{VA}).
\end{equation}
Then, we project the previous expression onto the tangential plane. Taking into account that the magnetic field has dominating polarization in the tangential plane provides
 \begin{equation}\label{H_VA=B_SA}
   \textbf{H}_{VA}=\frac{\textbf{B}_{VA}}{\mu_0}+ \hat{\textbf{u}}_{z}\times(\hat{\textbf{u}}_{z}\times\textbf{M}_{VA})\approx\frac{\textbf{B}_{SA}}{\mu_0}.
\end{equation}
This equation holds for the long-wavelength limit\cite{Silveirinha2009, Costa2011}. Thus, in order to calculate the correct values of the wave impedance (and Poynting vector) one can use \emph{volume averaged} numerical electric field $E_{VA}$, but \emph{surface averaged} numerical magnetic field $B_{SA}$
\begin{equation}\label{ZWcorrect}
   z_{W}=\frac{E_{VA}\mu_0}{Z_{0} B_{SA}}.
\end{equation}
We would like to remark that Eq.(\ref{ZWcorrect}) makes a bridge between our approach and that of papers with averaging field procedures \cite{Lerat2006_fieldsumm, Smith2006_fieldaver, Acher2007_fieldsumm}, where effective magnetic functions are obtained via volume averaging of $B$ fields, but surface averaging of $H$ fields.

Deriving effective permittivity and permeability from Eqs.~(\ref{EQ1}), (\ref{EQ2}) we find the MEP of the metamaterial. Accordingly reversing Eqs.~(\ref{EQ1}), (\ref{EQ2}) for the Bloch impedance (\ref{z_W}) we ends with the set of metamaterial WEP. Thus,
\begin{equation} \label{EP-material}
  \varepsilon_M=n/z_W, \quad \mu_M=nz_W ,
\end{equation}
and
\begin{equation} \label{EP-wave}
  \varepsilon_W=n/z_B, \quad \mu_W=nz_B .
\end{equation}
The latter should be equal to these given by the NRW method~\cite{Smith2002_SM}. We emphasize that determination of the propagation constants and impedances is straightforward, does not involve any inverse functions and is made on the basis of the same simulated fields for both wave and Bloch impedances.

We should mention a practical issue important for the implementation of the proposed approach. Computing fields by the finite-difference or finite-integral time-domain methods we should take into account a phase shift between the electrical and magnetic fields connected with the staggered Yee mesh. The electric and magnetic fields are calculated at different time moments shifted by $\Delta t/2$, where $\Delta t$ is the simulation time step. For the case of CST Microwave Studio, which we are using, the magnetic field phase is always shifted by $\Delta\phi=\omega\Delta t/2$, so we corrected the magnetic field values by corresponding phase factor $\exp(i \omega\Delta t/2)$.

\section{Specific examples of metamaterial structures} \label{CaseStudies}

\begin{figure}
    \includegraphics[width=\columnwidth,keepaspectratio]{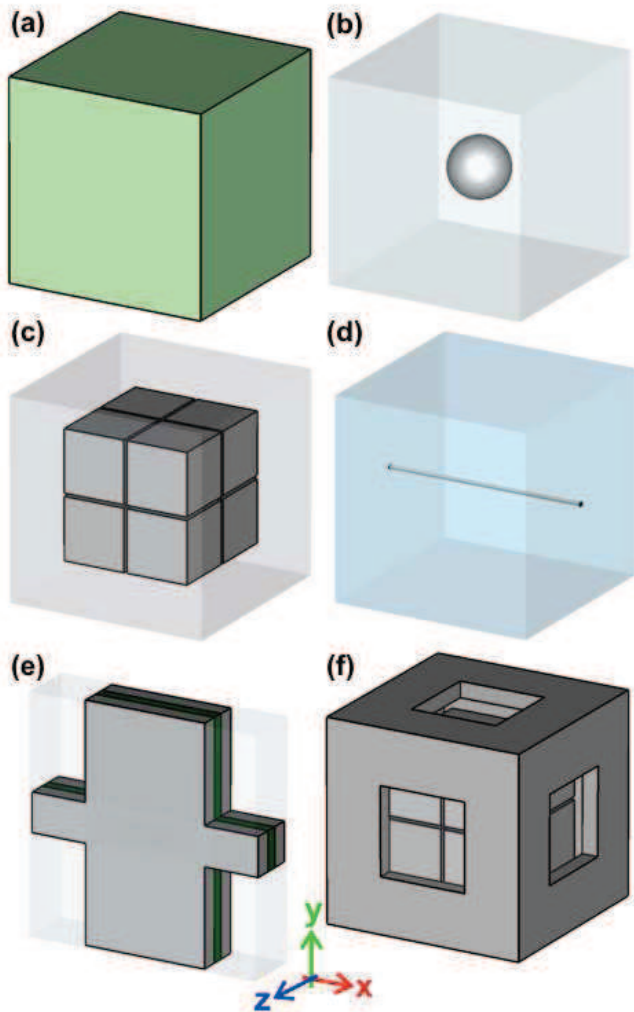}%
    \caption{\label{Sketches}(Color online). Sketches of the materials designs considered: homogeneous material (a), plasmonic nanospheres (b), split cube MM (c), wire medium (d), fishnet MM (e) and split cube in carcass MM (f).}
\end{figure}

We tested our approach on several examples, starting with the simplest ones. The unit cells sketches of the designs are shown in Fig.~\ref{Sketches}. We considered: (1)~homogeneous slab [see Fig.~\ref{Sketches}(a)]~-- two cases: lossless and Lorentz dispersion in $\varepsilon$ and $\mu$ with negative index of refraction, (2)~a set of the nanospheres with the plasmonic resonances [see Fig.~\ref{Sketches}(b)], (3)~split cubes MM that possess magnetic resonance and negative permeability [see Fig.~\ref{Sketches}(c)], (4)~wire medium that gives negative permittivity [see Fig.~\ref{Sketches}(d)], (5)~negative refractive index fishnet MM [see Fig.~\ref{Sketches}(e)], and (6)~split cube in carcass MM [see Fig.~\ref{Sketches}(f)]. In all cases, the MM slab consisted of 10 monolayers. For comparison, WEP for three-monolayers-thick slabs were calculated with the NRW method \cite{Smith2002_SM}.

\subsection{Homogeneous materials}

A slab of homogeneous material is the simplest object to test the retrieval approach, since the restored EPs can be compared with the exact values.

A homogeneous slab was artificially divided into 10 meta-atoms of the size $a_x=a_y=a_z=100~\mu$m. For the case of the homogeneous medium, the material and wave parameters are identical, so we should only compare the given constitutive parameters with the retrieved MEP.

For the homogeneous lossless slab with constant parameters: $\varepsilon=4$ and $\mu=1$ the EPs were in a perfect agreement with the theoretical permittivity and permeability (not shown). The relative retrieval error was less than 0.2\%, which can be attributed to numerical dispersion effect in finite-difference numerical simulations.

In another example, we consider the frequency dispersive permittivity and permeability described by the Lorentz model:
\begin{equation}
   \varepsilon(\omega)=\varepsilon_\infty+\varepsilon_{stat}\frac{\omega_{0e}^2}{\omega_{0e}^2-i\gamma_e\omega-\omega^2},
\end{equation}
\begin{equation}
   \mu(\omega)=\mu_\infty+\mu_{stat}\frac{\omega_{0m}^2}{\omega_{0m}^2-i\gamma_m\omega-\omega^2},
\end{equation}
where $\varepsilon_\infty=$1, $\varepsilon_{stat}=$1.7, $\omega_{0e}=2\pi \times 198 \times 10^9$~s$^{-1}$, $\gamma_{e}=2\pi \times 10^{10}$~s$^{-1}$, $\mu_\infty=$1, $\mu_{stat}=$1.3, $\omega_{0m}=2\pi \times 202 \times 10^9~$s$^{-1}$, $\gamma_{m}=2\pi \times 10^{10}$~s$^{-1}$.

\begin{figure}
    \includegraphics[width=\columnwidth,keepaspectratio]{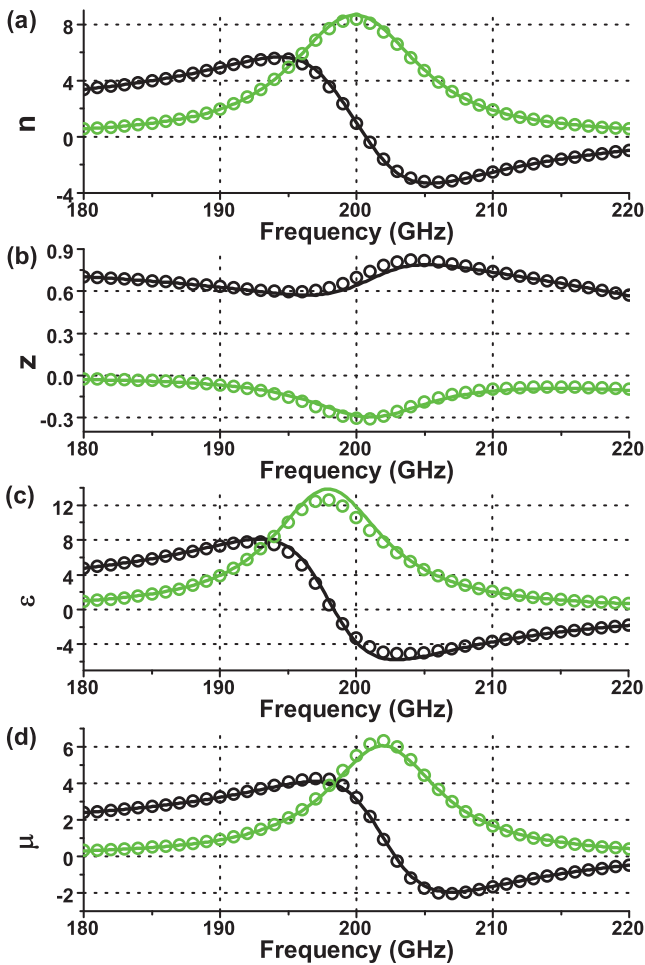}%
    \caption{\label{EPHomogeneous}(Color online). Retrieved effective parameters (circles) of the homogeneous medium with Lorentz dispersion in permittivity and permeability: refractive index (a), impedance (b), permittivity (c) and permeability (d), real (black) and imaginary (green/grey) parts. Results are compared with the original values (solid lines).}
\end{figure}

The restored effective parameters are in good correspondence with the original EPs [see Fig.~\ref{EPHomogeneous}]. The small differences are observed only in the resonant region around 200~THz where losses are high. The retrieval results in Fig.~\ref{EPHomogeneous} show that retrieving through the Bloch mode analysis is applicable to a range of  materials with or without losses with positive and negative $n, \varepsilon$ and $\mu$.

\subsection{Metamaterial composed of plasmonic nanospheres}

Metallic nanospheres possess plasmonic resonances. Being arranged in the regular structure, the nanospheres with a radius $r\ll\lambda$ make a MM. It is expected that the nanospheres MM should have the permittivity which is different from the host permittivity and its permeability should be close to 1, since the nanospheres are non-magnetic.

The silver nanospheres of the radius $r=30$~nm were placed in vacuum in the cubic array with the period $a_x=a_y=a_z=200$~nm. Silver was considered as the Drude metal \footnote[1]{The authors are aware that the permittivity of silver is not described correctly by the Drude formula in the optical range and experimentally measured data \cite{Palik_HOCS} should be used instead. However, we can adopt the simplified material dispersion for the sake of  testing the retrieval approach.} with the plasma frequency $\omega_p=1.37 \times 10^{16}$~s$^{-1}$ and collision frequency $\gamma_c=8.5 \times
10^{13}$~s$^{-1}$ (see Ref.~[\onlinecite{Dolling2006}]). The sketch of the design is shown in Fig.~\ref{Sketches}(b).

\begin{figure}
    \includegraphics[width=\columnwidth,keepaspectratio]{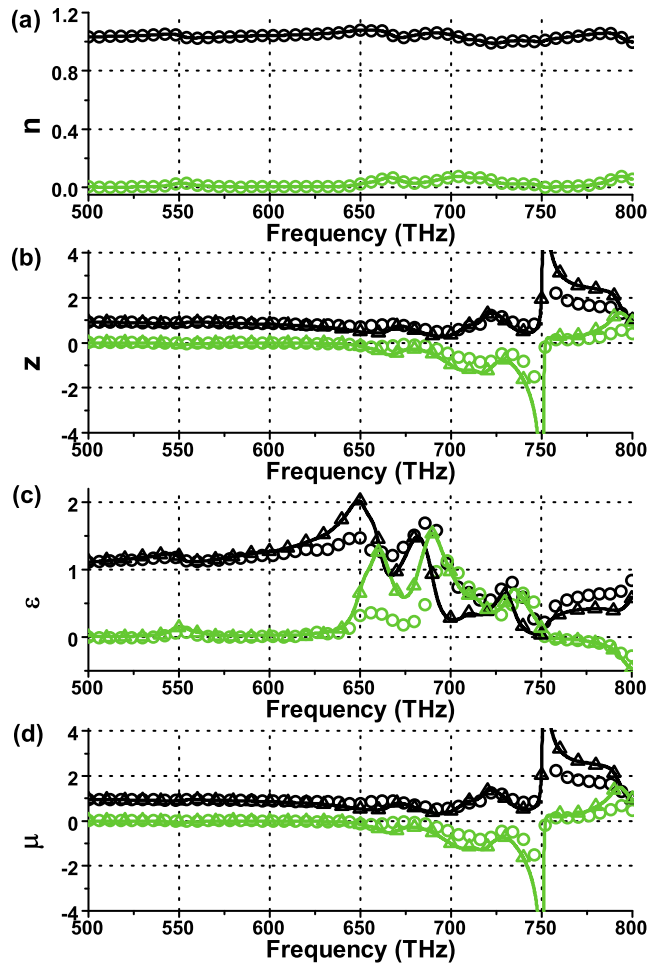}%
    \caption{\label{EPNanospheres}(Color online). Effective parameters of the MM consisting of plasmonic nanospheres: refractive index (a), impedance (b), permittivity (c) and permeability (d), real (black) and imaginary (green/grey) parts. Retrieved results by volume-averaged (circles) and surface-averaged (triangles) fields are compared with the NRW method (solid lines).}
\end{figure}

Effective refractive indices restored with the NRW method and our approaches are identical [see Fig.~\ref{EPNanospheres}(a)] as it was expected. Bloch impedance $z_B$, retrieved with the field surface averaging [see Fig.~\ref{EPNanospheres}(b), triangles] is identical to the one restored with the NRW method,  [Fig.~\ref{EPNanospheres}(b), solid lines].

There is a little difference between wave impedance $z_W$ [see Fig.~\ref{EPNanospheres}(b), circles]  and $z_B$ (triangles). They experience slight oscillations around the value of $z_W \simeq 1 + 0 i$. As a consequence of that both permittivities exhibit resonances around 660~THz, 690~THz and 730~THz [see Fig.~\ref{EPNanospheres}(c)], but of different strength. At the same frequencies, the magnetic permeability shows non-physical negative imaginary part, so-called antiresonance behavior that normally would correspond to the gain in the system. However, material EPs $\varepsilon_M$ and $\mu_M$, restored via the volume-averaged fields are free from the antiresonances on frequencies up to 700~THz. Small negative values of $\Im(\varepsilon_M)$ are due to the calculation errors with the staircase approximation of the spherical shapes.

The permeability $\Re(\mu)$, which is supposed to be around 1 since the nanospheres are non-magnetic, is indeed around 1 on frequencies up to 700~THz, but starts to oscillate on higher frequencies, especially at around 750~THz [see Fig.~\ref{EPNanospheres}(d)]. It looks as we have strong magnetism from the non-magnetic MM consisting of electric dipoles. In fact, at frequency 750~THz the condition for the first Bragg resonance is satisfied, so the MM cannot be considered as homogeneous and cannot be assigned with meaningful effective parameters ~\cite{Simovski2009_optspectrosc}.

\subsection{Split-cube metamaterial}

We choose a split cube MM as an example of a magnetic material with negative permeability in the infrared range~\cite{Andryieuski2009a,Andryieuski2009}. The sketch of the design, which is a 3D generalization of the symmetric split-ring resonator\cite{Penciu2008}, is shown in Fig.~\ref{Sketches}(c). The cubic unit cell of $a_x=a_y=a_z=250$~nm consists of the silver thin-wall structures (Drude metal) embedded in silica (permittivity 2.25). The geometrical parameters were taken the same as in the Ref.~[\onlinecite{Andryieuski2009}].

\begin{figure}
    \includegraphics[width=\columnwidth,keepaspectratio]{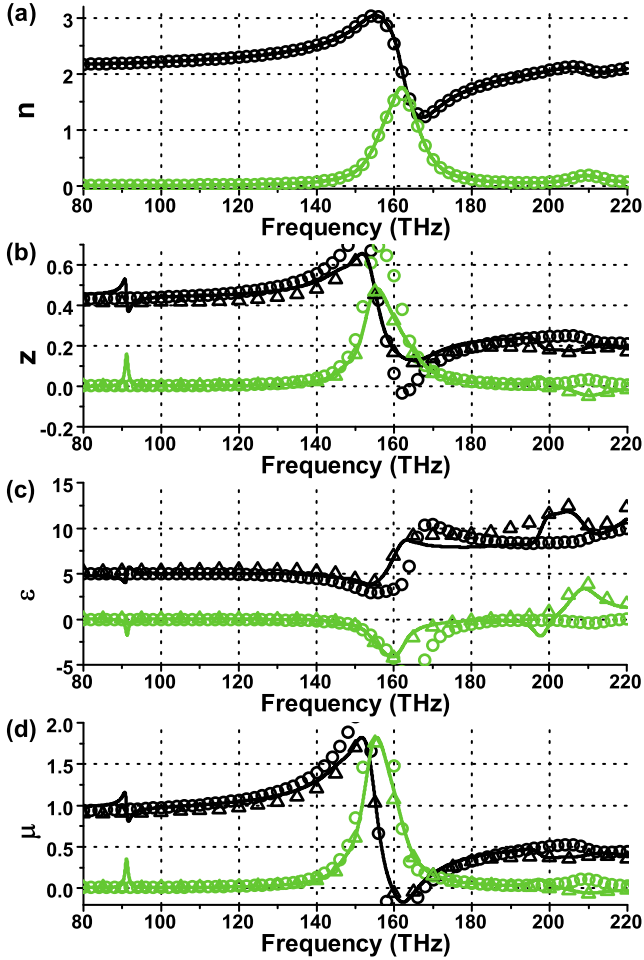}%
    \caption{\label{EPSCubes}(Color online). Effective parameters of the split cube magnetic MM: refractive index (a), impedance (b), permittivity (c) and permeability (d), real (black) and imaginary (green/grey) parts. Results by volume-averaged (circles) and surface averaged (triangles) approaches are compared with the NRW method (solid lines).}
\end{figure}

In the line with the previous cases, the refractive indices retrieved with different methods coincide, showing a resonance around 160~THz [see Fig.~\ref{EPSCubes}(a)]. Bloch and wave impedances exhibit strong resonance behavior in the area around 160 THz. A small peak in the impedance restored with the NRW method only at the frequency 91~THz appears at the Fabry-Perot resonance of the slab and is a numerical artifact intrinsic to the S-parameter method [see Fig.~\ref{EPSCubes}(b)]. The spurious peaks in the EPs due to Fabry-Perot resonances can be avoided with wave propagation methods as it was reported in Ref.~[\onlinecite{Andryieuski2009_WPRM}].

Effective parameters restored via surface and volume-averaged fields expose strong antiresonance behavior for the effective dielectric permittivity. Such behavior ordinary for WEP cannot be accepted in assigned MEP. The reasons for very similar appearance of effective parameters revealed by formulas (\ref{EQ2}),(\ref{z_W}) we assign to a strong magnetic resonance, which brings domination of magnetic field performance through $B_{SA}$ denominator and thus to formal equivalence of effective impedances. However, the full picture of failure of formula (\ref{z_W}) has yet to be understood.

\subsection{Wire-medium structure}

Wire medium \cite{rotman1962plasma} is a well-known example of the negative-permittivity MM. In the case of the square lattice of perfectly conducting wires in vacuum, when radius of the wires $r$ is much less than the unit cell size, $r\ll a$, an analytical expression for the effective permittivity is given in Ref.\cite{belov2002dispersion}:
\begin{equation}
   \varepsilon_{eff}(\omega)=1-\frac{2\pi c^2}{a^2\omega^2(\log\frac{a}{2\pi r}+0.5275)}.
\end{equation}

\begin{figure}
    \includegraphics[width=\columnwidth,keepaspectratio]{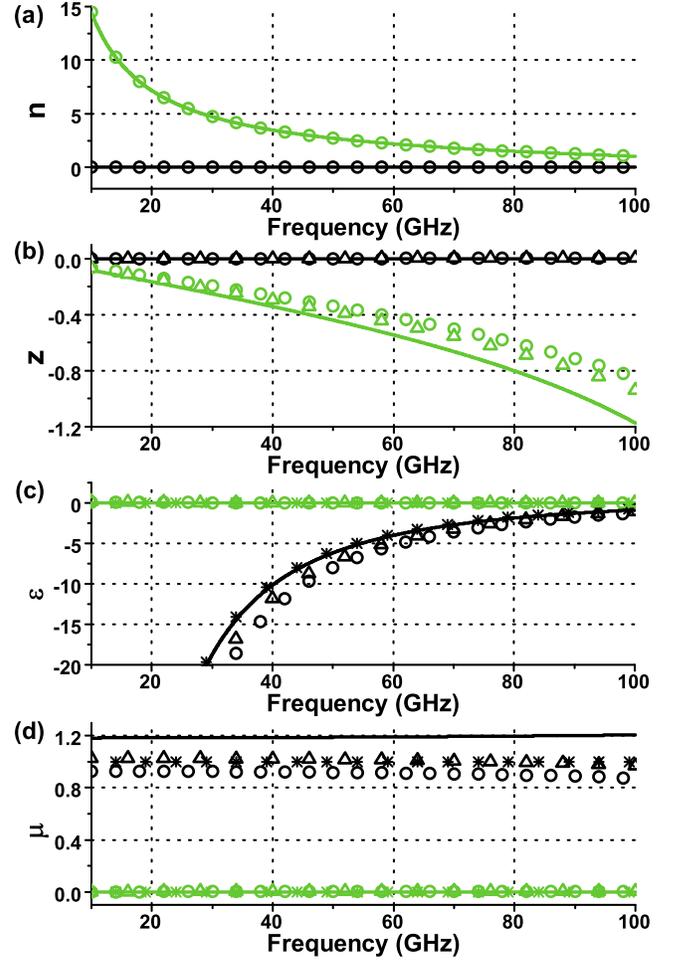}%
    \caption{\label{EPWM}(Color online). Effective parameters of the wire medium: refractive index (a), impedance (b), permittivity (c) and permeability (d), real (black) and imaginary (green/grey) parts. Results by the volume-averaged (circles) and surface averaged (triangles) approaches and NRW method (solid line) are compared with the analytical predictions (stars).}
\end{figure}

We simulated the $r=5~\mu$m-radius wires made from the perfect electric conductor arranged in a square lattice with $a_x=a_y=500~\mu$m in vacuum [see the sketch in Fig.~\ref{Sketches}(d)]. Comparison of the retrieved and analytical EPs is presented in Fig.~\ref{EPWM}.

Due to the rectangular spatial discretization of the round-shaped wires in the simulations we see the difference in the effective impedances retrieved through the field averaging procedure (both of them!) and the NRW method. It causes deviations in effective permittivities. Permittivity retrieved with the NRW method is closer to the analytical results [see Fig.~\ref{EPWM}(c)]. What concerns permeability, the NRW method retrieves paramagnetic $\Re(\mu_W)\approx1.2$ [see Fig.~\ref{EPWM}(d)], while the wire medium is expected to be a non-magnetic MM. Within the field-averaging approach the retrieved $\mu_W$ perfectly coincides with the theoretical prediction, while $\mu_M$ seems to be more sensitive for the staircase approximation errors.

We should note that because we study wave propagation perpendicular to the wires no any spatial dispersion effect showed up during the restoration, and results are physically sensible.

\subsection{Fishnet metamaterial}

The fishnet MM\cite{Dolling2006} is one of the most promising negative-index metamaterials for the optical and infrared regions. It consists of the metallic double wires extending in the $x-$ and $y-$ directions [see the sketch in Fig.~\ref{Sketches}(e)].

We use the geometrical and material parameters of the fishnet MM from Ref.~[\onlinecite{Rockstuhl2008_fishnet_BM}] except adjusting the period in $z-$direction to $a_z=150$~nm. The unit cell transverse sizes are $a_x=a_y=600$~nm. Silver layers (silver treated as the Drude metal) of the thickness 45~nm are separated with the $Mg$$F_2$ dielectric with refractive index $n=1.38$ and thickness 30~nm. This metal-dielectric-metal sandwich is placed in vacuum.

\begin{figure}
    \includegraphics[width=\columnwidth,keepaspectratio]{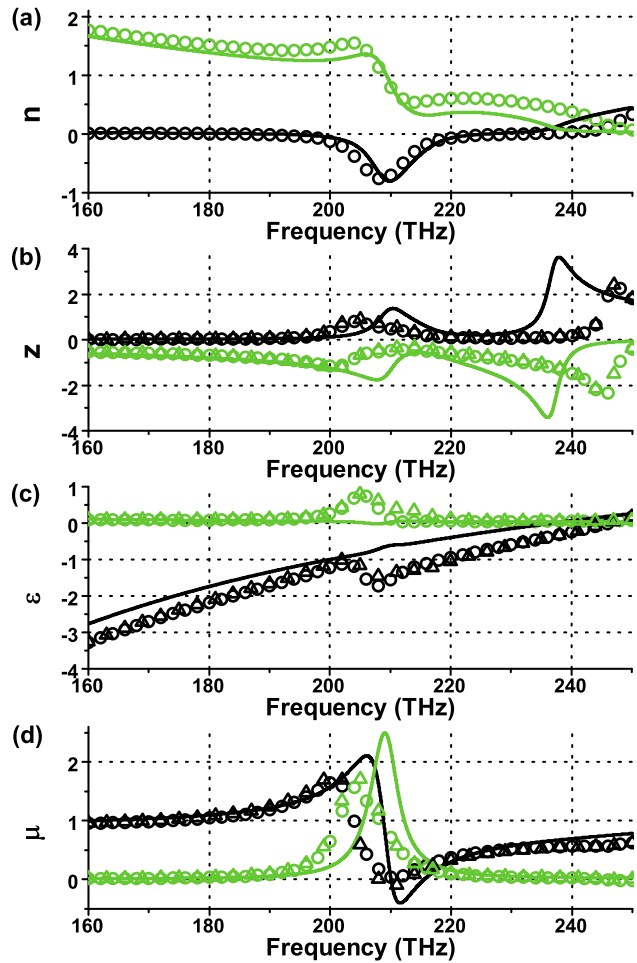}%
    \caption{\label{EPFishnet}(Color online). Effective parameters of the fishnet negative-index MM: refractive index (a), impedance (b), permittivity (c) and permeability (d), real (black) and imaginary (green/grey) parts. Results by volume-averaged (circles) and surface averaged (triangles) approaches are compared with the NRW method (solid lines).}
\end{figure}

The refractive indices retrieved with our approach and the NRW method are slightly different [see Fig.~\ref{EPFishnet}(a)]. This is not surprising since the NRW method is applied to a three monolayers-thick slab. It is well known that the thin-slab effective refractive index of the fishnet converges slowly to the bulk values with the increase of the slab thickness~\cite{Belov2010_FNhomog,Zhou2009}. Our approach based on field propagation in 10 layers gives the  refractive index close to its bulk values. Bloch and wave impedances are different as well [see Fig.~\ref{EPFishnet}(b)]. We also expect that the NRM results would converge to ours if ten layers will be considered.

Effective parameters obtained by both types of field averaging are quite close to each other. The feature of the fishnet behavior is the negative index region free from the anti-resonances both in $\varepsilon$ and $\mu$. The NRW results exhibit hardly visible anti-resonance for $\Im(\varepsilon)$, which is corrected via field-averaging procedures.

\subsection{Split-cube-in-carcass metamaterial}

A fishnet MM is an example of a medium with a negative refractive index. To check that we can assign effective parameters, which will not show any antiresonances, we consider another negative-index metamaterial with strong spatial dispersion, namely split cube in carcass\cite{Andryieuski2009_WPRM,Andryieuski2009} [see the sketch in Fig.~\ref{Sketches}(f)]. Its remarkable property is extreme fast convergence of parameters such that its effective refractive index is the same for the 1-layer thick slab and for the bulk MM represented by the infinite number of layers. However, as was shown in Ref.~[\onlinecite{Menzel2010}], even being 3D cubic symmetric by design, split cube in carcass is anisotropic in the resonant region.

The cubic unit cell of $a_x=a_y=a_z=250$~nm [Fig.~\ref{Sketches}(f)] consists of the silver split cube (the same as in [Fig.~\ref{Sketches}(c)] nested in the silver carcass, which is a kind of 3D wire medium. The metallic structures are embedded in silica.

\begin{figure}
    \includegraphics[width=\columnwidth,keepaspectratio]{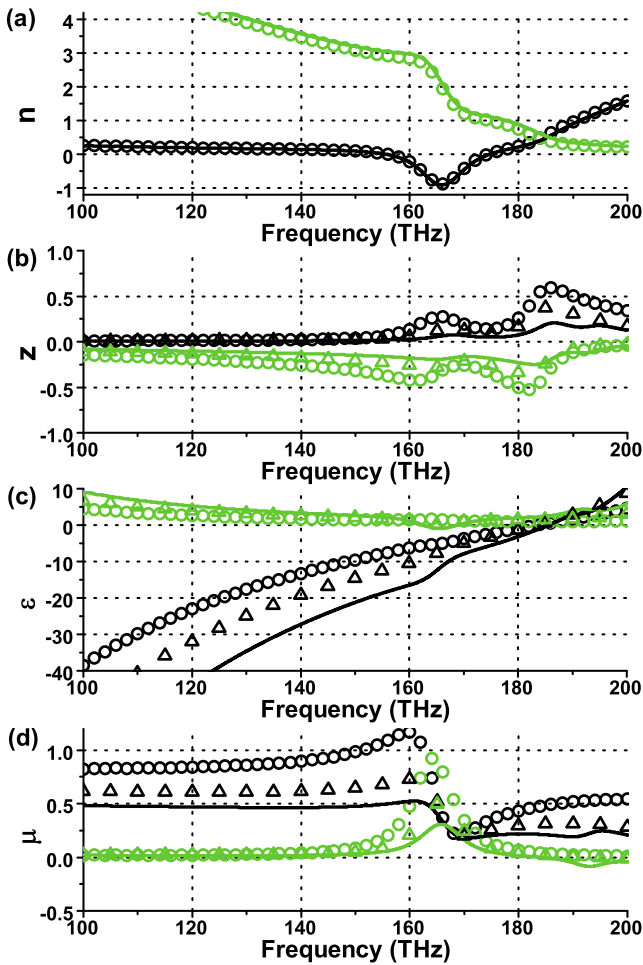}%
    \caption{\label{EPSCiC}(Color online). Effective parameters of the split cube in carcass negative index MM: refractive index (a), impedance (b), permittivity (c) and permeability (d), real (black) and imaginary (green/grey) parts. Results by the volume-averaged (circles) and surface averaged (triangles) approaches are compared with the NRW method (solid lines).}
\end{figure}

As the effective refractive index of the split cube in carcass does not depend on the slab thickness, it is not surprising that the NRW method and our approach give results coinciding much better than for the fishnet [see Fig.~\ref{EPSCiC}(a)]. Nevertheless, effective impedances, and therefore permittivities and permeabilities provided by all three approaches are different [see Figs.~\ref{EPSCiC}(b,c,d)].

We should also note that in both cases in the frequency ranges beyond the resonances the volume averaging approach produces physically sound results. As an illustration we note that diamagnetism observed in the $\Re(\mu_W)$ does not remain in the $\Re(\mu_M)$, which is close to conventional 1 below the resonant region.

\section{Discussion and Conclusions}
\label{Conclusions}

We have suggested a novel approach for the extraction of effective parameters of metamaterials based on the study of dispersion properties of the Bloch waves propagating in quasi-periodic structured materials. In all the cases with single-mode propagation our approach provides solid results for the effective refractive indices, which can be attributed to the bulk refractive indices  of the metamaterials irrespectively of their anisotropy and spatial dispersion. Our spectral analysis approach is able to retrieve refractive indices for a wide range of materials and structure geometries, which can be lossy or lossless, dispersive, possess negative permittivity, permeability and refractive index values. The method is simple and unambiguous, free from the "branch" and Fabry-Perot problems, which are the issues for the reflection/transmission based NRW method. The results provided by the NRW method are identical to the results obtained by our method in all considered cases except for the case of the fishnet MM, where EPs experience poor convergence to the bulk values. The single-mode propagation of a MM can be checked during the retrieval process from the fields mismatch monitoring procedure.

The spectral analysis serves as a platform for further advance in retrieving EPs. Impedance retrieving is very sensitive to the conditions of restoration and can lead either to WEP or MEP. Employing surface averaged fields of the dominating Bloch mode, we obtain WEP, which are nearly identical for those retrieved by the NRW method, but free from spurious resonances appearing from the Fabry-Perot effects in slabs. All what is needed for the MEP retrieval accordingly to Ref.[\onlinecite{Simovski2009_optspectrosc, Simovski2011}] is the volume averaging of the electric and magnetic fields over the unit cell. Both retrievals (wave and material EPs) are performed within a single computational cycle, because fields on the unit cells entrance facets or in its volumes  are available, and they can be exported from Maxwell's solver arrays. The approach works for MM slabs with thicknesses at least 3-4 monolayers. Our approach adequately reveals the typical non-magnetic behavior of metamaterials away from the resonance regions, which is problematic for the NRW method. Therefore, we anticipate that the proposed approach will become a useful tool for the characterization of both wave and material effective properties of MMs.

\begin{figure}
    \includegraphics[width=\columnwidth,keepaspectratio]{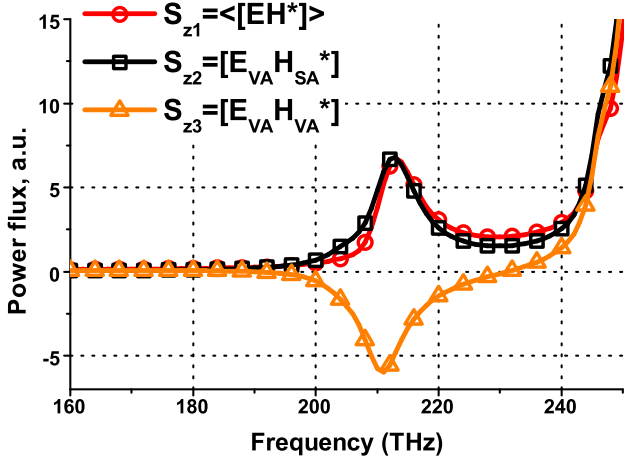}
    \caption{\label{Poynting}(Color online) $z$-component of the Poynting vector of the fishnet negative-index MM: volume averaged Poynting vector (red line with circles), correctly defined Poynting vector for the fundamental Bloch harmonic (black line with squares) and flux calculated through the volume averaged electric and magnetic fields of the fundamental Bloch harmonic (orange line with triangles).}
\end{figure}

It should be noted that the magnetic microfields returned by Maxwells solvers are $\textbf{b}/\mu_{0}$-fields, while the volume averaged magnetic field $\textbf{H}_{VA}$ must be used. Possible implications of ignoring this fact can be illustrated through the Poynting vector calculations, Fig.(\ref{Poynting}). Here the fishnet structure from the Subsection 3D is used. Poynting vectors are calculated accordingly to three formulas:
\begin{equation} \label{all Sz-impedances}
   \begin{array}{l} {\displaystyle
       S_{z1}=\Re(\int_{V}[\textbf{e}\times \textbf{h}^{\ast}]dV),
       } \\ {\displaystyle
       S_{z2}=\Re[\textbf{E}_{VA}\times \textbf{H}_{SA}^{\ast}],
       } \\ {\displaystyle
       S_{z3}=\Re[\textbf{E}_{VA}\times \textbf{H}_{VA}^{\ast}].
    } \end{array}
\end{equation}

 Straightforward calculations of the Poynting vector give us the negative $z$-component $S_{z3}$ (see orange line with triangles in Fig.(\ref{Poynting}), which means that vectors $\textbf{k}$ and $\textbf{S}$ are parallel in the negative index domain. This is exactly what can happen if the wrong formulation of the Poynting vector through vector $\textbf{b}$ is used as it is pointed in Refs.[\onlinecite{Costa2011,Silveirinha2009}]. The consequences of this is not only the wrong direction of the flux, but also the negative value of the $\Re(z)$, because flux and impedance are connected through the expression

\begin{equation} \label{Sz-impedance}
   \begin{array}{l} {\displaystyle
       S_{z3}=\Re(\textbf{e}_{z}[\textbf{E}_{VA}\times \textbf{H}^{*}_{VA}]) =
   } \\ {\displaystyle \quad
       = \Re(E_{VA} H^{*}_{VA}) = Z_{0} \Re(z_{W})|  H_{VA}|^{2}.
   } \end{array}
\end{equation}

  However, employment of the volume averaged electric and surface averaged magnetic fields improves the situation (black line with squares). The Poynting vector $S_{z2}$ calculated through them is very close to the averaged microscopic flux $S_{z1}$ (red line with circles). Such calculations confirm the fact that on the grid level, microfields $\textbf{b}$ and $\textbf{h}$ differ only by a constant. But fields averaged over a macrovolume bear principle differences.

\begin{figure}
    \includegraphics[width=\columnwidth,keepaspectratio]{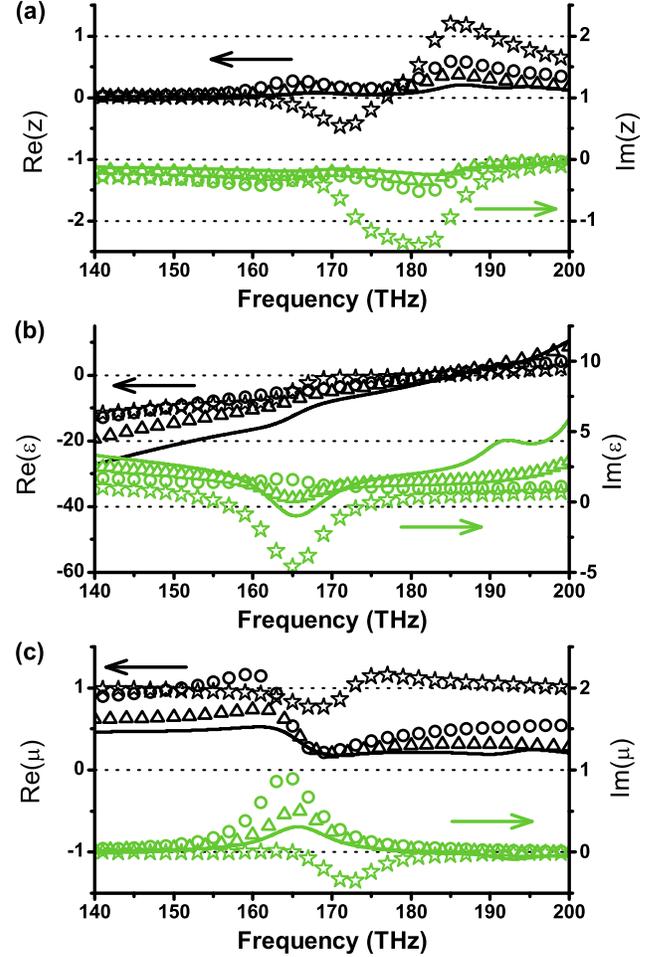}
    \caption{\label{Comparizon_z}(Color online) Effective parameters of the split cube in carcass negative index MM: impedance (a), permittivity (b) and permeability (c), real (black) and imaginary (green/grey) parts. Results are obtained by formula (\ref{ZWcorrect}) (circles), (\ref{z_B}) (triangles), (\ref{z_W}) (stars) and the NRW method (solid lines)approaches .}
\end{figure}

The most intriguing part is the direct comparison between effective parameters restored with formulas (\ref{z_B}), (\ref{z_W}) and (\ref{ZWcorrect}). In Fig.(\ref{Comparizon_z}) we plot results for three different cases of impedance restoration and include also the NRW restoration data. In fact, the volume-averaged fields provide the incorrect result (stars), with negative $\Re(z)$ and double anti-resonances in $\Im(\varepsilon)$ and $\Im(\mu)$. The situation is improved when the surface-averaged (transverse-averaged) fields are taken (triangles) instead of bulk fields in concordance with the finding in \cite{Silveirinha2007a}. There is still one faint "attempt" of an antiresonance with decreasing of $\Im(\varepsilon)$. And there is completely no antiresonance, when using the formula (\ref{H_VA=B_SA}). The corresponding curves are designated by circles in Fig.(\ref{Comparizon_z}).

Unfortunately this approach cannot be accepted as universal retrieving method, because in some cases (see Split Cube case in Section 3C) it fails. More deep analysis in the failure of formula (\ref{ZWcorrect}) is needed, but it lies beyond the scope of this paper.

We should admit that a direct extension of our approach for the experimental characterization of MMs in the optical range is challenging, since there are no such small electric and magnetic field detectors that could be placed inside the MM unit cell without noticeable influence on its functionality. Nevertheless, as the radio and microwave frequency range, it is possible to record the fields at the spatial points inside the metamaterial~\cite{Sukhorukov:2009-3716:OE}, enabling the direct application of our approach.

\begin{acknowledgments}

The authors thank C.R. Simovski, A.V. Vinogradov, and C. Menzel for stimulating discussions of the Bloch and wave impedances. The special thanks are given to the unknown reviewer and M. Silveirinha for drawing our attention to the nature of computed fields and useful advice on the magnetic field correction. A.A. and A.V.L. acknowledge a financial support from the Danish Research Council for Technology and Production Sciences via the NIMbus and GraTer (11-116991) projects. A.A.S. and Y.S.K acknowledge a support from the Ministry of Education and Science of Russian Federation (Russia) and Australian Research Council (Australia).

\end{acknowledgments}

\bibliographystyle{osa3AuAll}
\bibliography{Andryieuski}

\end{document}